# 3D modelling of strain concentration due to PCI within the fuel code ALCYONE

J. Sercombe, J. Julien, F. Michel, B. Michel, E. Fédérici.

*CEA, DEN, DEC/SESC, F-13108 Saint-Paul-lez-Durance, France.*
*Tel: 0442253072, Fax: 0442252949 , Email: jerome.sercombe@cea.fr*

**Abstract –** In this paper, the 3D scheme of the fuel code ALCYONE is applied to a database consisting of more than 50 base irradiations and ramp tests performed on rods with $UO_2$, MOX or Cr-doped $UO_2$ fuels and Zy4 or M5® cladding tubes with burn-ups up to 70 GWd/tU. The ability of the 3D scheme to predict the behaviour of a single fuel pellet – cladding element situated at the maximum Linear Heat Rate (LHR) during ramp testing is demonstrated by comparing the following experimental and calculated data: residual clad diameter after base irradiation and ramp test, height of inter-pellet and mid-pellet ridges after base irradiation and ramp test, dish filling after ramp test. In the second part of the paper, 3D simulations that catch the stress – strain localization at the triple point are presented. It is shown that the stress-strain concentration occurs only at the end of the power transient and is quickly relaxed by pellet and clad creep.

## I. INTRODUCTION

ALCYONE is the multi-dimensional fuel code co-developed in the PLEIADES platform [1] for non-accidental [2][3] and accidental situations [4] by the CEA, EDF and AREVA. It incorporates three different calculation schemes which describe the thermo-mechanical behaviour of a complete fuel rod (1D axisymmetric) or that of a single fuel pellet fragment and overlying cladding (2D plane strain or 3D representation) during irradiation in commercial Light Water Reactors (LWR) or power ramps in experimental reactors.

PCI failures usually begin at the so-called triple point (axial location: Inter-Pellet plane, circumferential location: in front of a radial pellet crack, radial location: inner clad wall) where the stresses and strains are maximum during a power transient [5]. The detailed description of the pellet geometry (dishing, chamfer) together with sophisticated material laws for pellet cracking and creep incorporated in the 3D model [6] makes it therefore a powerful tool to study stress and strain localization during Pellet Cladding Interaction (PCI). A prerequisite is that the 3D model gives a realistic representation of clad residual strains after base irradiation and power ramps.

In the first part of this paper, the application of the 3D scheme to a database consisting of more than 50 base irradiations and ramp tests performed on rods with $UO_2$, MOX or Cr-doped $UO_2$ fuels and Zy4 or M5® cladding tubes with burn-ups up to 70 GWd/tU is described. The ability of the 3D scheme to predict the behaviour of a single fuel pellet – cladding element situated at the maximum Linear Heat Rate (LHR) during ramp testing is demonstrated by comparing the following experimental and calculated data: residual clad diameter after base irradiation and ramp test, height of inter-pellet and mid-pellet ridges after base irradiation and ramp test, dish filling after ramp test.

In the second part of the paper, 3D simulations that catch the stress – strain localization at the triple point are presented.

## II. DESCRIPTION OF ALCYONE

ALCYONE is a multi-dimensional application which consists of four different schemes concerned with :

- the complete fuel rod discretized in axial segments (1D), see Figure 1,
- half of the pellet and the overlying cladding (2Drz), see Figure 2 where the Mid-Pellet (MP) plane is situated at the top of the mesh,
- one quarter of a pellet fragment and associated cladding (3D), see Figure 2, where the MP plane is situated at the top of the mesh,
- the mid-pellet plane of the 3D pellet scheme (2Dr$\theta$), see Figure 2.

The different schemes use the same Finite Element (FE) code CAST3M to solve the thermo-mechanical pellet-gap-cladding problem and share the same physical material models at each node or integration points of the FE mesh. This makes the comparison of simulated results from one scheme to another possible with no dependency on the constitutive models.

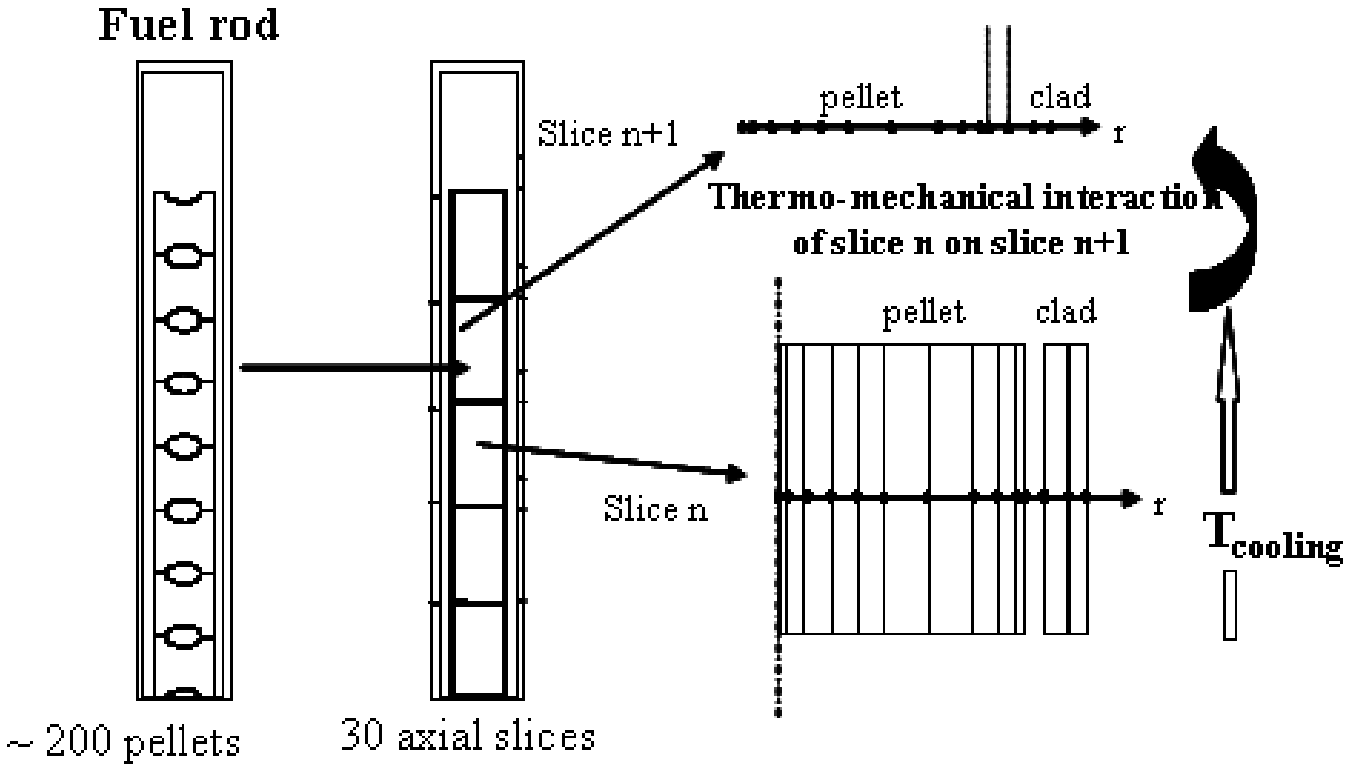

Fig 1. 1D scheme of ALCYONE.

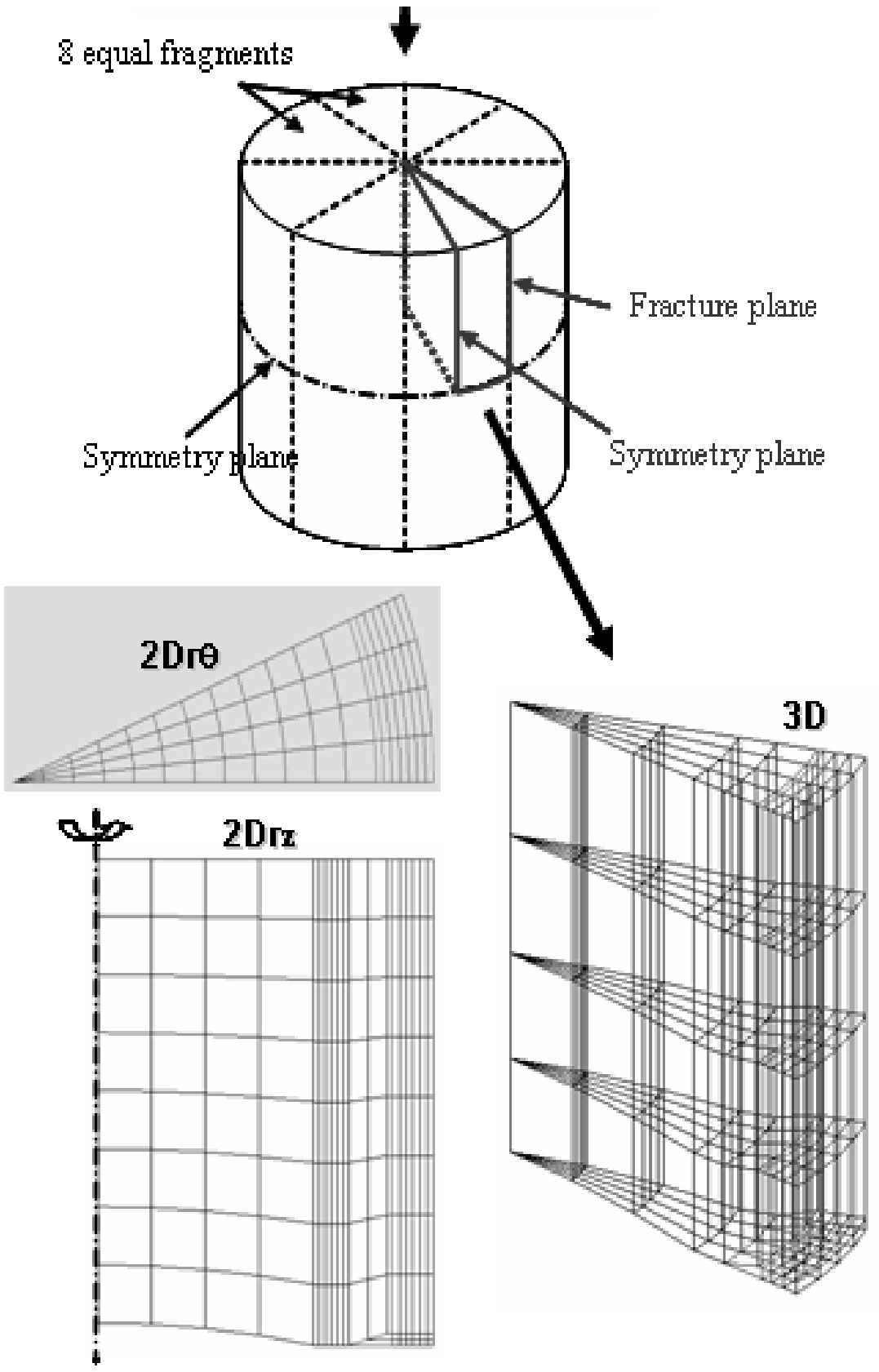

Fig 2. 2D and 3D schemes of ALCYONE

The main phenomena considered in the simulations of base irradiations and power ramps are :

- Irradiation creep, thermal creep and plasticity of the cladding during base irradiation and power ramp testing.
- Creep and cracking of the fuel pellet during base irradiation and power ramp testing.
- Evolution of the thermal and mechanical properties of the materials with temperature, porosity and burnup.
- Coupled thermo-mechanical analysis of the fuel pellet – gap – cladding system (the gap size depends on the deformation of the rodlet and on the Fission Gas release).
- Generation of, diffusion of Fissions Gas in intra-granular, inter-granular and connected pores, release of FG in the plenum (internal pressure update).
- Pellet densification, pellet FG-induced swelling (stress-dependent).
- Relocation of pellet fragments after pellet cladding contact.

Note that fuel cracking and fuel relocation are treated as separate phenomena with distinct kinetics.

### *II.A. The 1D scheme*

Performing 1D simulations of the complete fuel rod during base irradiation and power ramp testing is a preliminary step before using the other schemes of ALCYONE. 1D simulations provide information on the FG release kinetics and hence on the internal pressure in the fuel rod, axial temperature and pressure distribution in the coolant, cladding elongation, … ALCYONE 1D predictions in terms of profilometry, gas release or elongation are validated on experimental data concerning 80 base irradiations and 30 ramp tests performed on rods with $UO_2$, MOX or Cr-doped $UO_2$ fuels and Zy4 or M5® cladding tubes with mean burnups up to 80 GWd/tU [7]. In ALCYONE 1D, relocation of pellet fragments after pellet cladding contact derives from an empirical relation based on 3D simulations. Creep and radial cracking of the fuel pellet are described by a creep-damage material law [8].

### *II.B. The 3D scheme*

In the 3D FE model of ALCYONE, only one quarter of a single pellet fragment and the overlying piece of cladding (see Figure 3) are meshed. The size of the pellet fragment is consistent with post-irradiation examinations performed on PWR pellets after 2 or 3 cycles of base irradiation [9], which show the existence of, on average, 10 pellet fragments in the circumferential direction. For symmetry reasons, it is assumed in the model that the pellet is initially fragmented in 8 identical pieces. The pellet-cladding system is meshed with more or less 1500 solid quadrangles with 8 nodes. The pellet description accounts for the geometrical particularities of the fuel element (dishing, chamfer, central void, …).

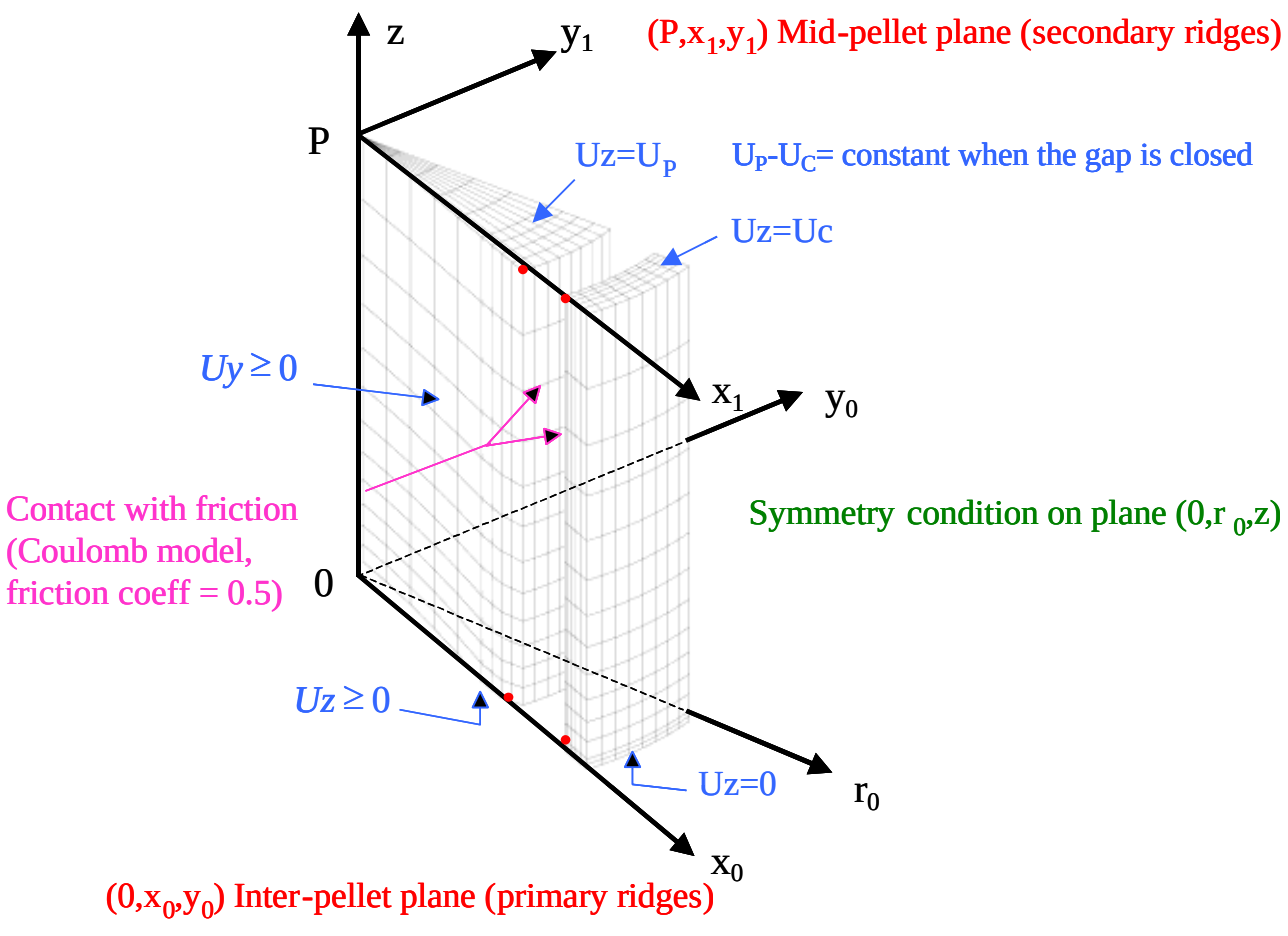

Fig. 3. Mesh and boundary conditions in the ALCYONE 3D simulations.

The boundary conditions considered in the 3D calculations are shown in Figure 3. They account for the geometrical symmetries of the problem and for the pellet-cladding and pellet-pellet interactions. At the inter-pellet plane (plane $0x_0y_0$ in Figure 3), unilateral contact conditions are prescribed ($Uz \geq 0$). The mechanical reaction of the fissile column above and under the meshed fragment is represented by a kinematics relation between the pellet and the cladding mid-planes (plane $Px_1y_1$ in Figure 3). This mid-plane locking condition is applied only when the pellet cladding gap is closed at least at one point. Pellet-pellet inter-penetration along the fracture plane $0x_0z$ is forbidden by the unilateral contact condition $Uy \geq 0$. Concerning loading conditions, the internal pressure (gas pressure) is applied to the cladding inner surface and to the pellet fragment outer surface. The external pressure (water pressure) is applied to the cladding outer surface.

The phenomena and models used in the 3D scheme are identical to those of the 1D scheme, except for the relocation which stems directly from the thermo-mechanical behaviour of the pellet-cladding fragment and for fuel cracking which is described by an elasto-plastic softening law with the r, θ and z axis as the principal cracking directions [10]. Closing of cracks is taken into account during reverse loading. Concerning the pellet-cladding interface, unilateral contact is assessed and a "Coulomb" model is introduced to simulate friction-slip or adherence. The latter is of primary importance with respect to stress and strain concentration in the cladding. A "burnup-dependent" coefficient of friction is used in the simulations which is consistent with the evolution of pellet-clad interface during irradiation [9].

## III. 3D VALIDATION PROCESS

### *III.A. The database*

The database used for the 3D validation process consists in about 50 ramp tests performed on 2 to 6 cycles base irradiated rods with $UO_2$, MOX or Cr-doped $UO_2$ fuels and Zy4 or M5® cladding tubes at maximum Linear Heat Rates between 395 and 610 W/cm and with holding periods between 0 and 12 hours. Most of the ramp tests have been performed in the OSIRIS reactor at the CEA Saclay in France [5]. Post-Irradiation Examinations (PIE) include measures of the residual diameters and height of Inter-Pellet (IP) and Mid-Pellet (MP) ridges after base irradiation and power ramp tests, estimations by optical microscopy of dish filling of the pellet and of the number of radial cracks by after ramp tests. This database provides a wide variation of loading conditions which in turn can lead to important differences in terms of IP or MP ridges formation. The geometry of the pellets is however fairly constant with a Height over Diameter ratio of about 1.5-1.6 and includes dishings, 6 mm in diameter and 0.3 mm in depth. About 40% of the pellets have no chamfer.

The 3D calculations are performed as follows : for each fuel rodlet ramp tested, the position of the maximum LHR is determined. The power history, external cladding temperature and coolant pressure are then extracted from 1D simulation results. Simulations are then performed with the set of parameters determined for the 1D complete fuel rod calculations (in particular those concerning the fission gas model). From the deformation of the cladding at the end of the calculation, the out-of-ridge residual diameter, the height of IP and MP ridges are determined [5], see Figure 4.

From the deformation of the meshed pellet dishing, residual dish filling can be assessed. The 3D calculations presented in this paper use a burnup-dependent creep model for $UO_2$ in order to account for the reduction in creep rate due to fission products in the matrix [3].

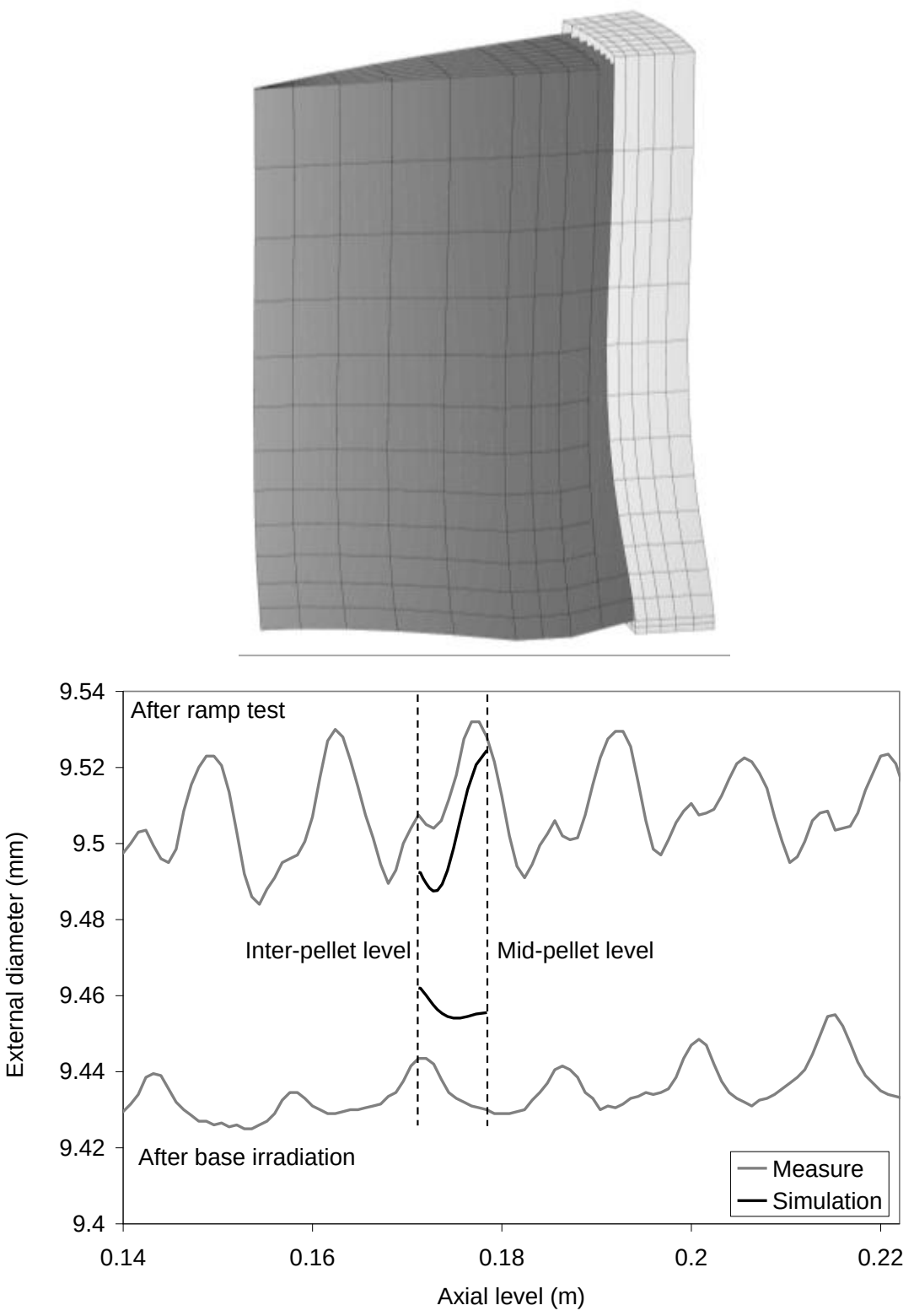

Fig. 4. Illustration of the radial displacement of the pellet and cladding after ramp testing showing the IP (bottom) and MP (top) ridges.

*III.B. Profilometry at the end of base irradiation*

The calculated out-of-ridge residual diameters of the cladding after base irradiation are compared in Figure 5 to the experimental measures (mean value from measures performed on 8 generatrices) . The plain line represents the equality between calculated and measured diameters.

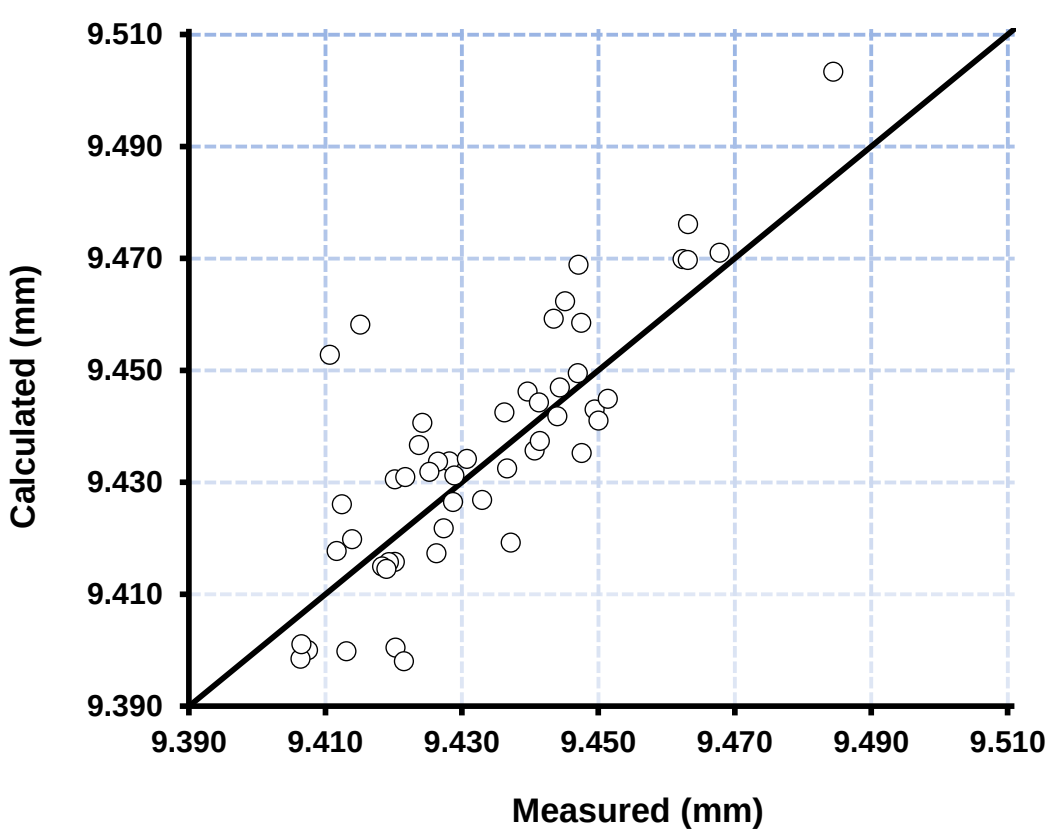

Fig. 5. Calculated and measured out-of-ridge residual diameters of the cladding after base irradiation.

On average, the 3D simulations tend to give out-of-ridge residual diameters overestimated by 4 microns. These results are considered satisfactory when compared to the accuracy of the measure, e.g., ± 5 microns.

The height of IP ridges at the end of base irradiation are compared in Figure 6 to experimental measures (mean value estimated from 7 successive IP).

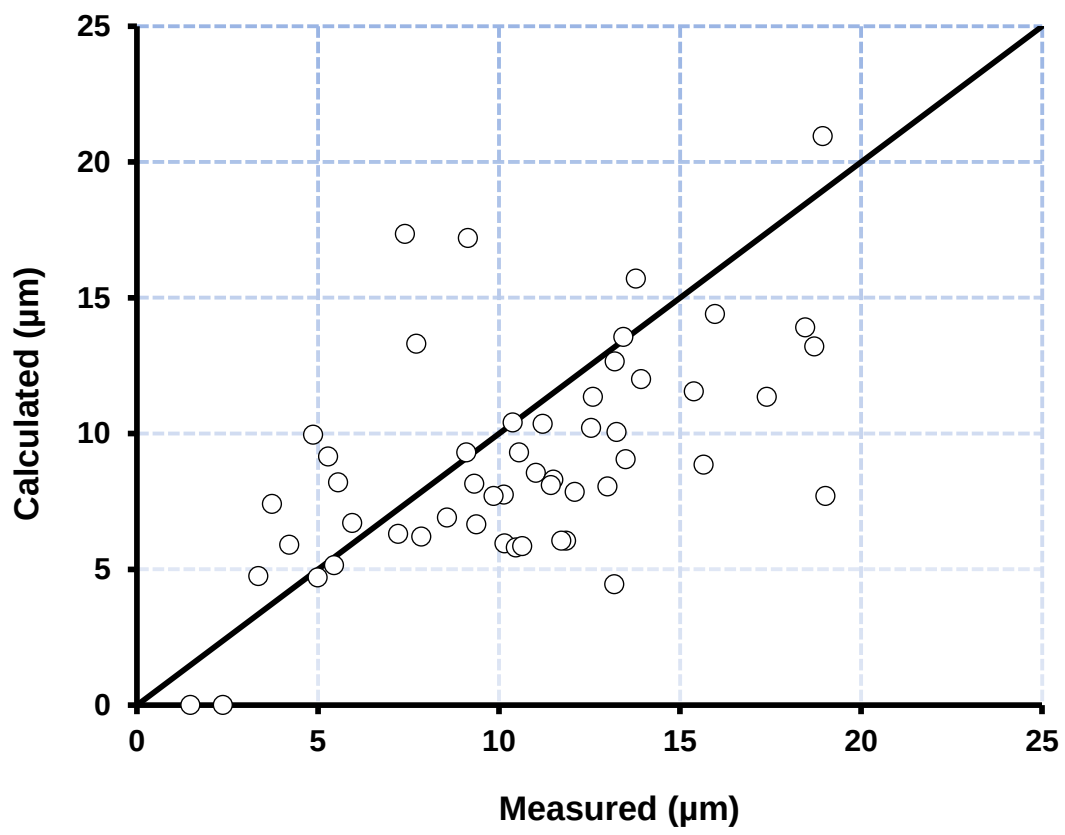

Fig. 6. Calculated and measured height of Inter-Pellet ridges at the end of base irradiation

On average, the 3D simulations underestimate the height of IP ridges by 1.6 microns. These results fall within the experimental scatter associated with the measured height of IP ridges which is of ± 2.2 microns on 7 successive IP.

*III.C. Profilometry after ramp tests*

The calculated and measured mid-pellet diameter increase during ramp testing (mean value from measures performed on 8 generatrices) are compared in Figure 7.

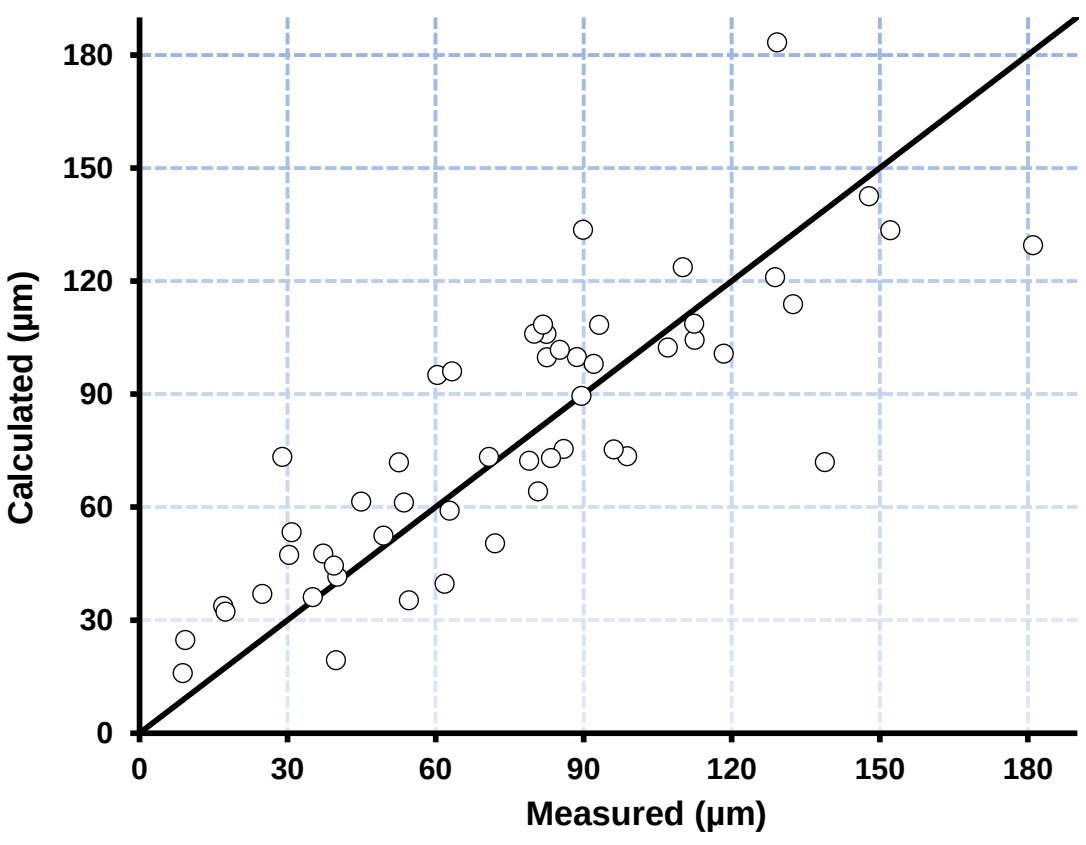

Fig. 7. Calculated and measured mid-pellet diameter increase during ramp testing.

The 3D simulations lead to a small overestimation of 4.7 microns of the mid-pellet diameter increase during ramp test. The agreement between simulated and measured diameters is very good. Calculated MP and IP ridges heights after ramp testing are compared in Figures 8 and 9 to experimental measures (mean value estimated from 7 successive pellets situated at the axial position of the maximum LHR).

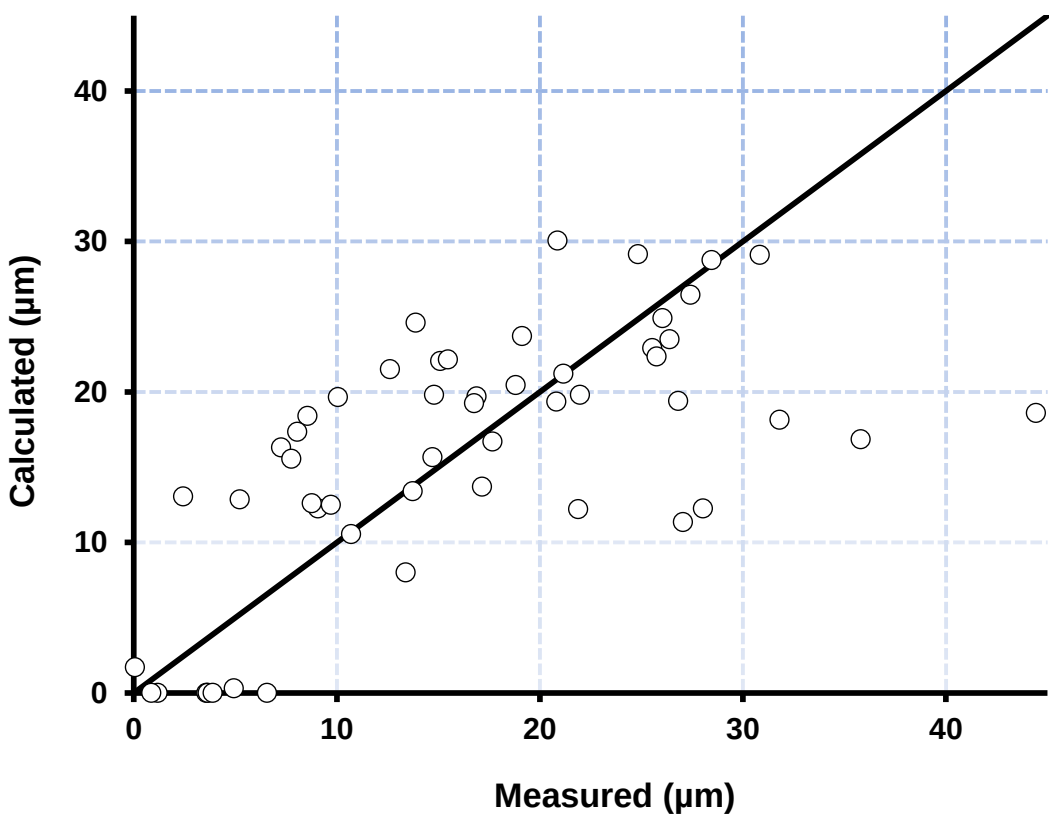

Fig. 8. Calculated and measured height of Mid-Pellet ridges at the end of the ramp test

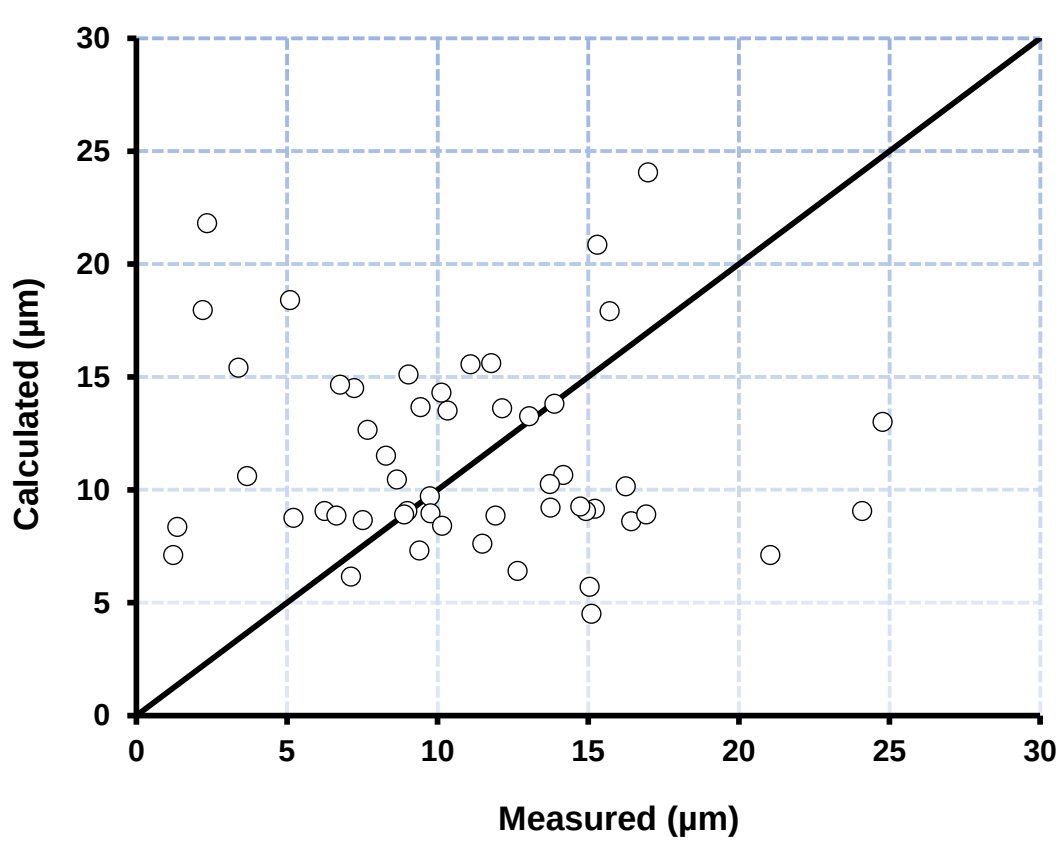

Fig. 9. Calculated and measured height of Inter-Pellet ridges at the end of the ramp test

On average, the height of MP and IP ridges after ramp testing are estimated by ALCYONE 3D within less than 1 micron. These results fall within the experimental scatter associated with the measured heights of IP and MP ridges which are of ± 3 microns on 7 successive pellets. A more important dispersion of data appears in Figures 8 and 9 due to the fact that the calculated IP and MP ridges after ramp tests include the contribution of base irradiation loading with already some differences with experimental measures. For some fuel rods, the over- or underestimations observed after ramp tests stems from differences already there after base irradiation.

The experimental and calculated diameter increase at IP level are plotted in Figure 10. On average, ALCYONE 3D tends to overestimate the diameter increase at IP level by 6 microns. The error is very close to the dispersion observed on the 7 pellets situated at the maximum LHR (close to 5 microns).

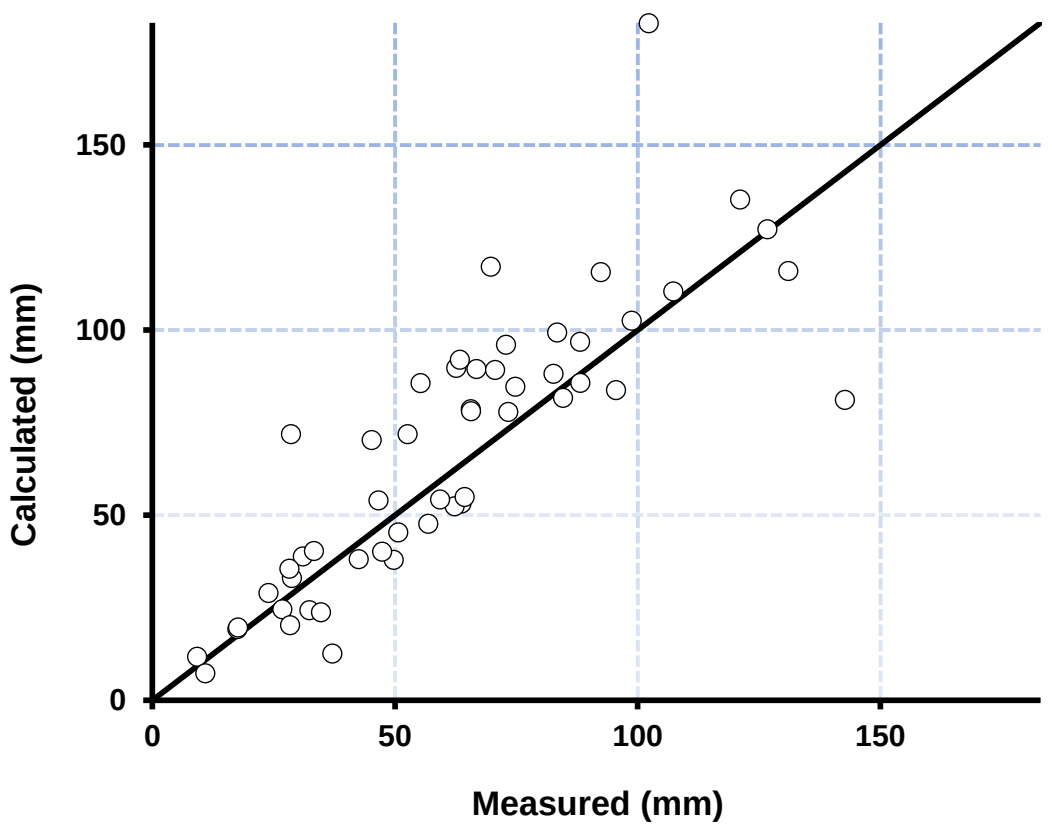

Fig. 10. Calculated and measured inter-pellet diameter increase during ramp testing.

This precise estimation of IP diameter increase is a major progress with respect to the previous validation of the fuel code ALCYONE 3D [2]. The behavior of the fuel pellet at IP level is rather complex since it results from a compromise between the pellet thermal and gas-induced swelling which tends to increase the diameter, and the filling up of the dishing due to $UO_2$ creep at high temperature which tends to reduce the diameter expansion. A dual experimental - 3D modeling analysis of dish filling during power ramps showed that creep laws established on non-irradiated $UO_2$ samples tend to overestimate systematically dish filling rates. To correct this behavior, a burnup-dependent creep law has been proposed and calibrated by comparing experimental and simulated dish fillings on the ramps with no holding time [3].

Figure 11 gives an overview of the calculated dish fillings (residual) compared to experimental estimations. All the ramps of the database are included, the ones with a long holding period give the points on the right part of Figure 11 (residual dish filling close to 80%). Figure 11 shows that the calculated dish fillings are very well estimated with the burnup dependent creep law. Since dish filling has a major impact on clad diameter increase at IP level during ramp test, this explains the improved prediction by ALCYONE 3D of this quantity as show in Figure 10. The correct estimation of the clad deformation at IP level is of great importance for the prediction of clad failure.

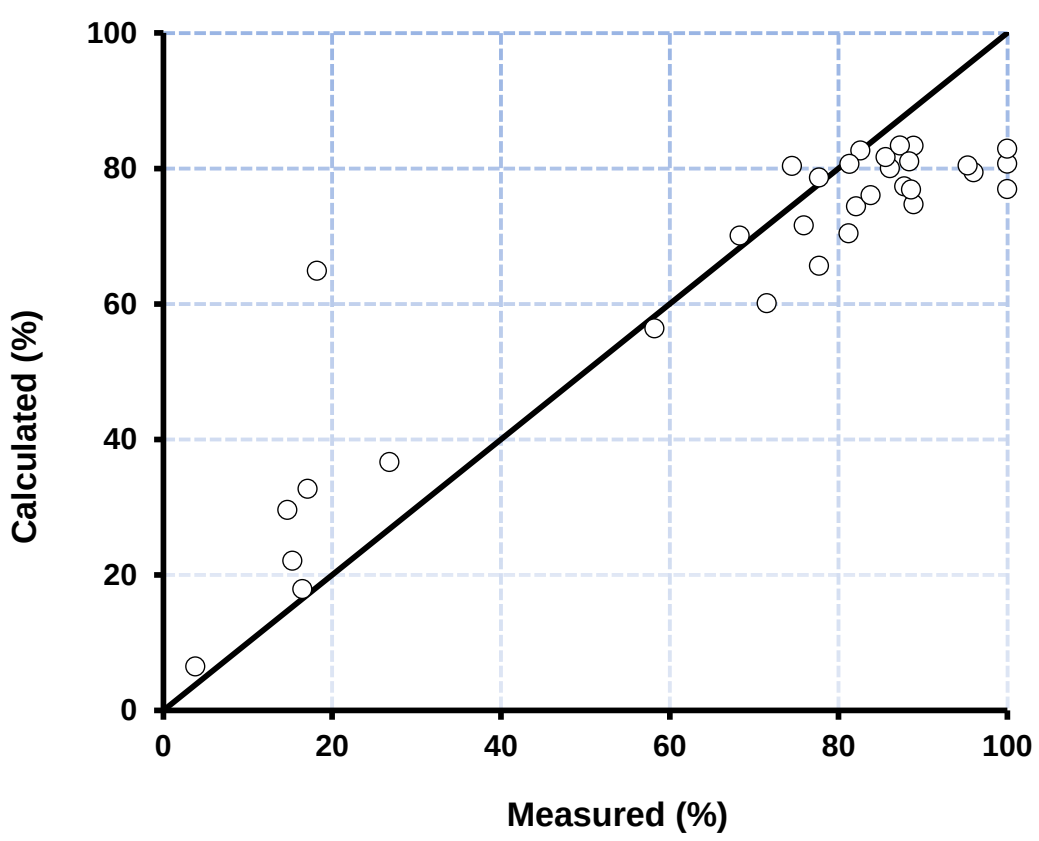

Fig. 11. Calculated and measured dish filling at the end of ramp tests.

## IV. STRESS CONCENTRATION

Clad failures by I-SCC (Iodine Stress Corrosion Cracking) usually begin at the so-called triple point (axial location: Inter-Pellet plane, circumferential location: in front of a radial pellet crack, radial location: inner clad wall) where the stresses and strains are maximum during a power transient [11], see Figure 12. Catching the stress or strain concentration at the triple point requires a 3D simulation with a mesh size consistent with the stress discontinuities at the pellet-clad interface [12]. In the circumferential direction, the (r,θ) shear stress and contact pressure drop to zero in front of the pellet crack opening. The crack opening gives therefore the size of the discontinuity in the circumferential direction. During a power transient, based on analytical models [13], it is typically 10-30 microns wide. The mesh size in this direction should therefore be close to 1 micron. In the axial direction, considering only chamfered pellets, the chamfer height gives the typical dimension. A mesh size of 20 microns is therefore necessary to trigger the stress concentration in the axial direction.

To catch the stress localization at the triple point, a series of 3D simulations have been performed with the non uniform mesh given in Figure 13. The total number of elements (5000) is about 3 times that of the mesh of Figure 3 used for the validation process and has not therefore increased greatly. The big difference lies in the number of potential contact – friction relationship at the pellet-clad interface which has increased by a factor 10 due to the refinement of the mesh near the triple point.

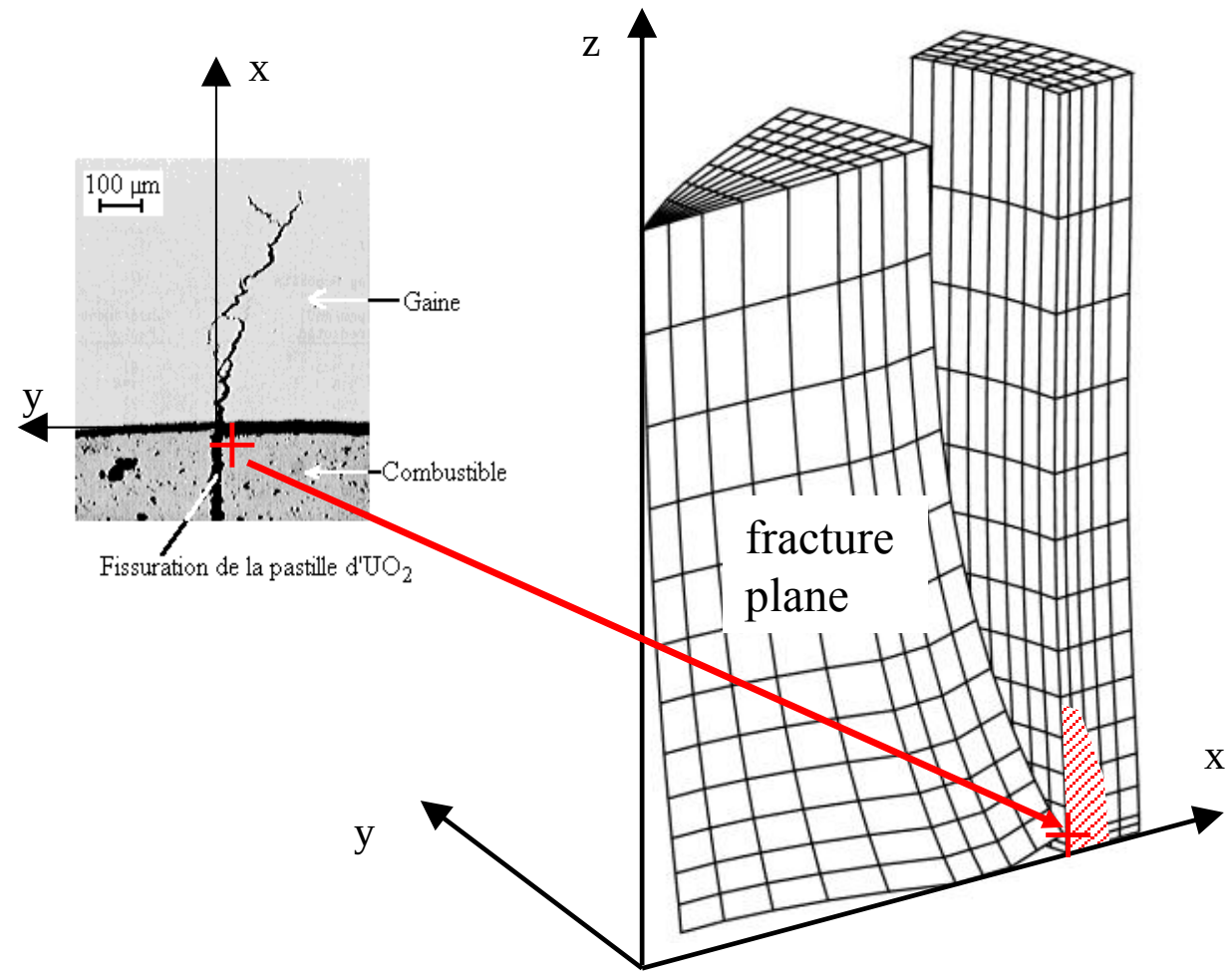

Fig. 12. Illustration showing the position of the triple point in the 3D simulation.

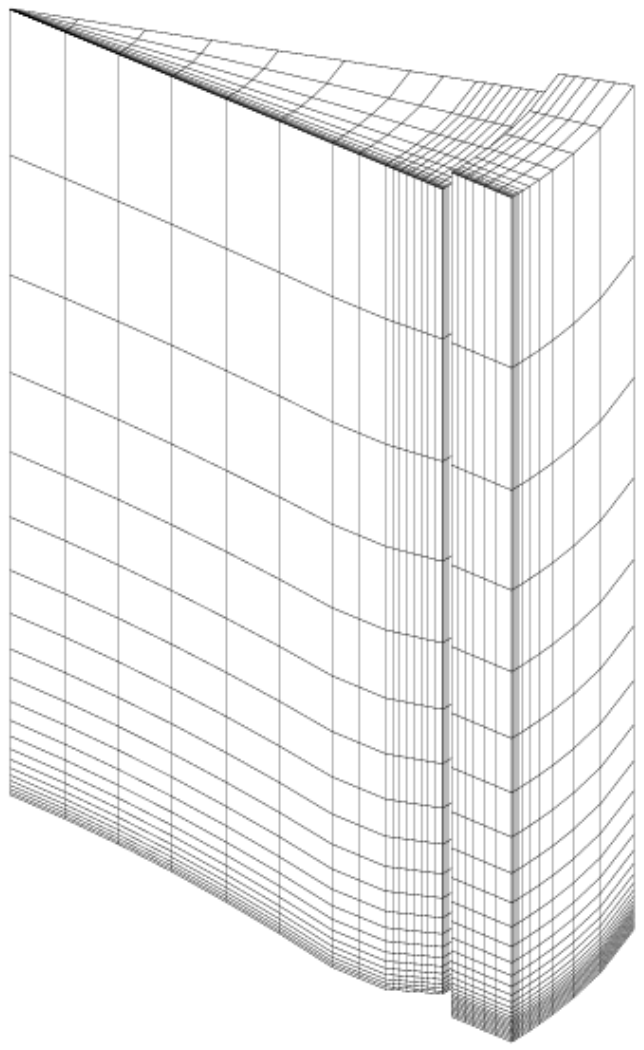

Fig. 13. Mesh used in the 3D simulations to catch the stress-strain concentration at the triple point.

The evolution of the hoop stress on the inner clad wall at different times during a power ramp that led to the failure of a $UO_2$-Zy4 rod after a few minutes is presented in Figure 14. Note that a very good agreement between measured and calculated residual clad diameters and ridges after base irradiation and ramp test has been obtained for this fuel rod. The calculated dish filling (around 30%) was also consistent with the post-ramp metallography.

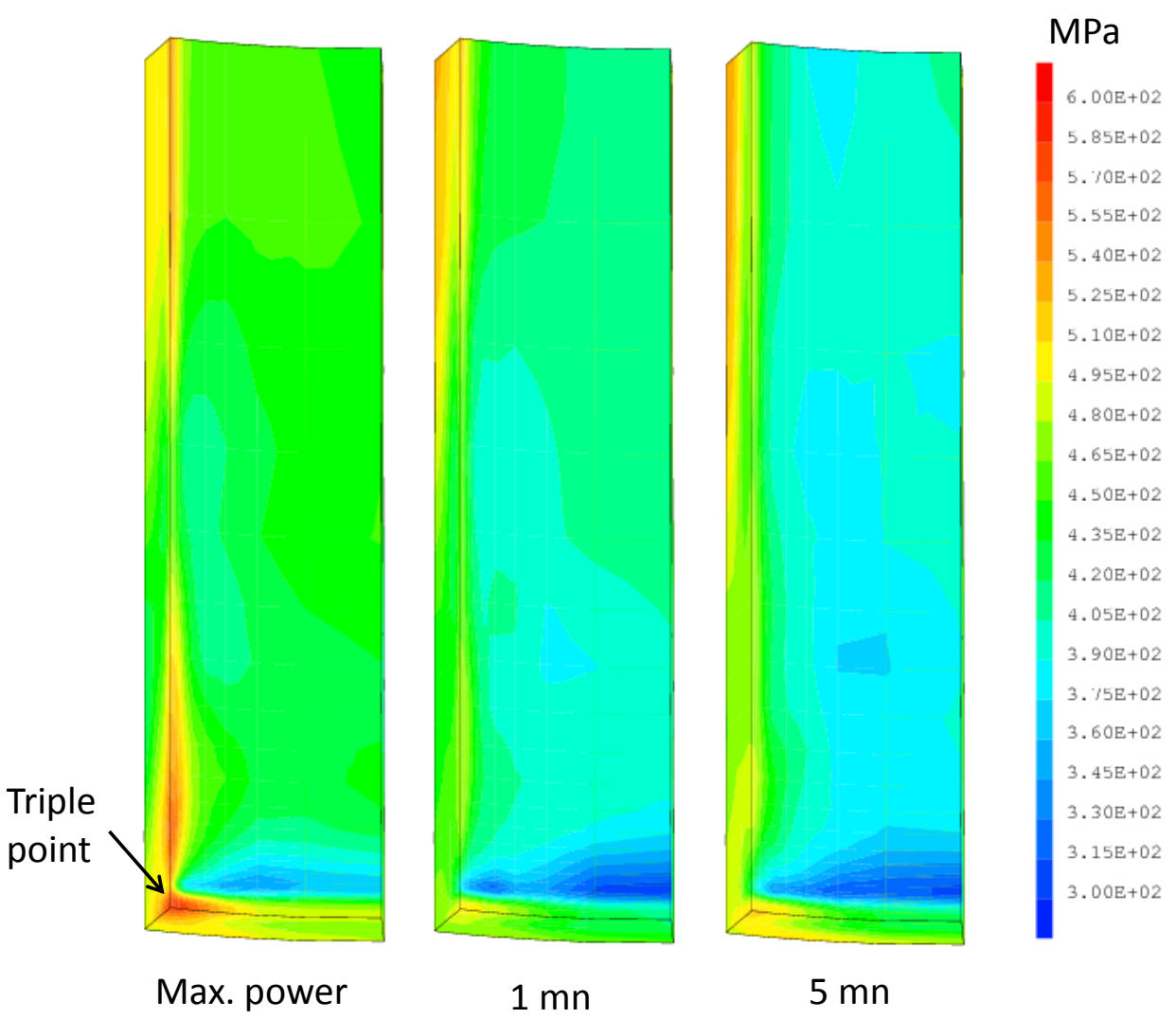

Fig. 14. Evolution of the hoop stress during a power ramp that led to the failure of the rod (max. power, after 1 mn at max. power, after 5 mn at max. power).

The localization of the hoop stresses at the triple point is clearly seen in Figure 14 at maximum power at the end of the transient. It is situated at the exact IP level and limited axially by the pellet-clad contact in front of the chamfer (materialized by the blue color in Figure 14). This is due to the very high contact pressure (radial stresses close to 250 MPa) that takes place in front of the chamfer. The maximum hoop stress reaches almost 600 MPa. The hoop stress gradient across the clad thickness is also particularly important in front of the triple point. It exceeds 400 MPa/mm. After a few minutes at maximum power, the hoop stress distribution is already very different from that obtained at the end of the transient. Clad creep enhanced on the inner wall due to the higher temperatures leads to some reduction in the maximum hoop stress which is now close to 500 MPa. The hoop stress gradient disminishes considerably becoming close to 100 MPa/mm after 5 minutes.

The evolution of the inelastic (plastic + creep) hoop strains on the inner clad wall during the power ramp is illustrated in Figure 15. The strain localization at the triple point, consistent with the stress distribution, is also seen at the end of the transient (max. power). The maximum strain on the inner clad wall reaches 0.6%. The strain gradient across the clad thickness at the triple point is vey high, reaching almost 1%/mm. Five minutes later, the situation is very different since the maximum inelastic hoop strains are obtained at mid-pellet level. This is due to the creep of the pellet which leads to a reduction of the prescribed strains at IP level. In consequence of the high height/diameter ratio of the pellets, the diameter increases at Mid-Pellet level leading to Mid-Pellet ridges as shown in Figure 16.

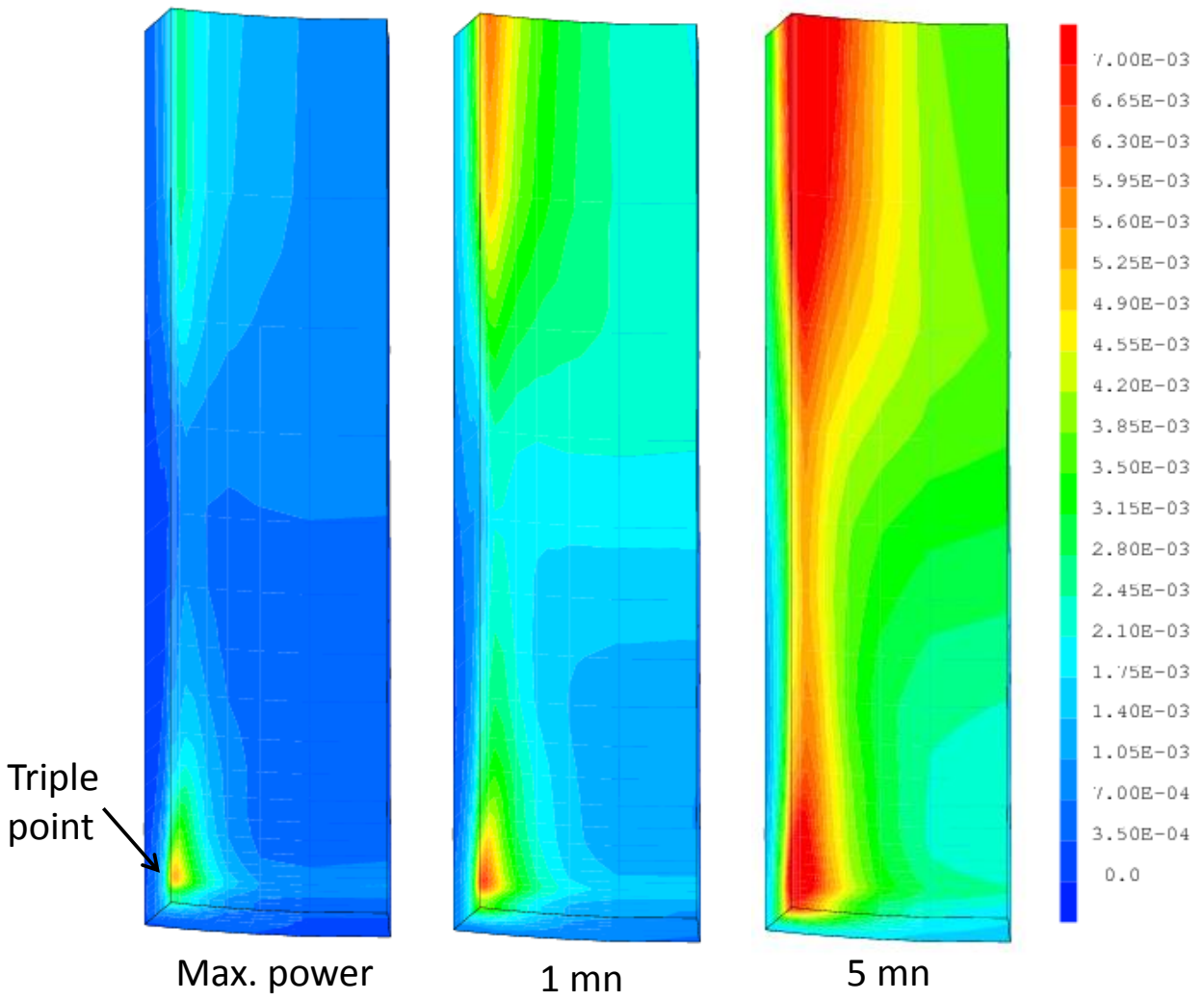

Fig. 15. Evolution of the inelastic strains (plastic + creep) during a power ramp that led to the failure of the rod (max. power, after 1 mn at max. power, after 5 mn at max. power).

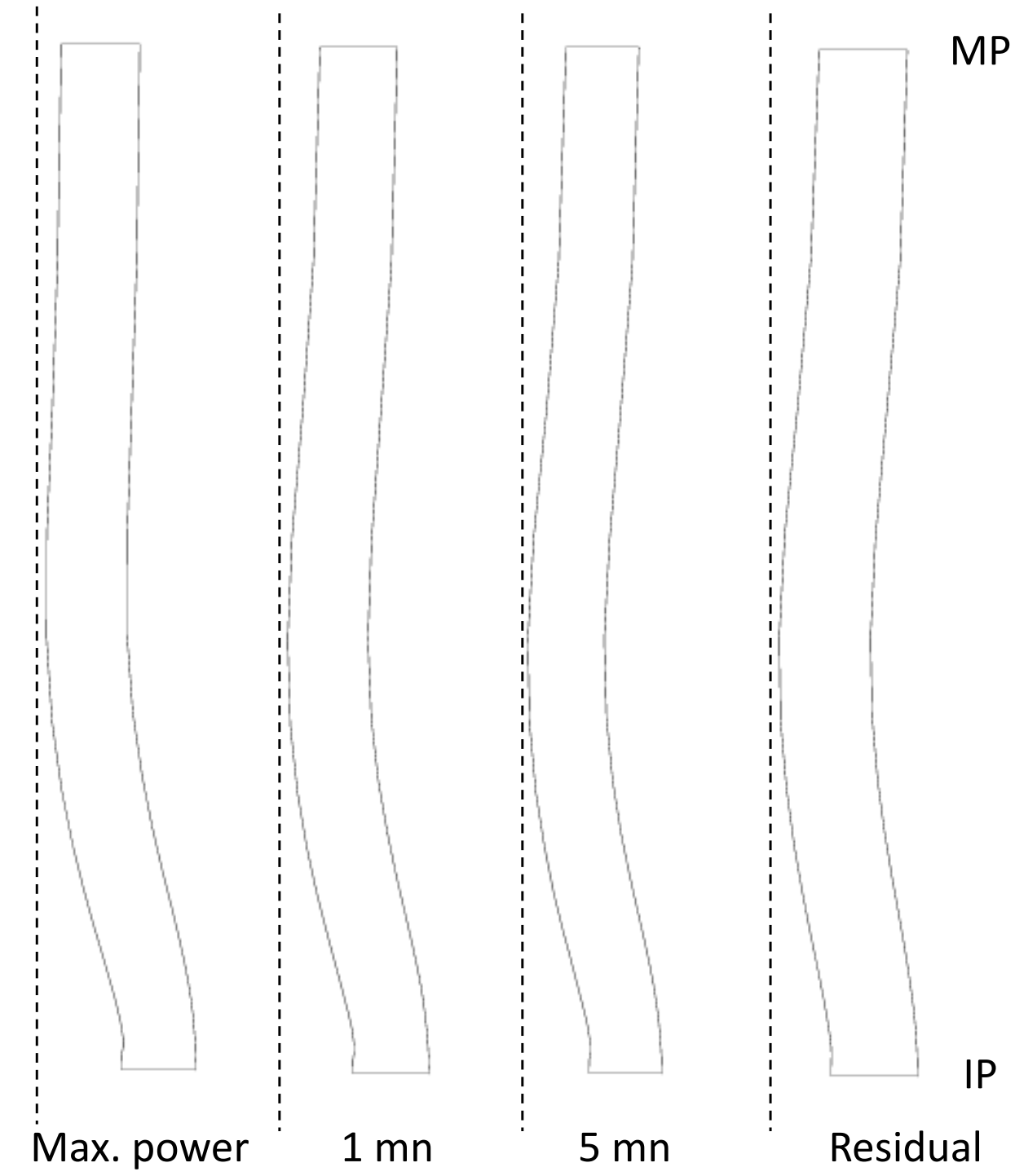

Fig. 16. Evolution of the clad deformation during a power ramp that led to the failure of the rod (max. power, after 1 mn at max. power, after 5 mn at max. power).

These Figures show that even if the rod presents mid-pellet ridges as high as the inter-pellet ridges at the end of a power ramp, the loading of the clad inner wall at the end of the transient is concentrated at the triple point where I-SCC is generally found to initiate.

## V. CONCLUSIONS

In this paper, 3D simulations of PCI during base irradiation and ramp tests have been presented. The 3D scheme of ALCYONE was applied to a large database consisting of more than 50 ramp tests performed on $UO_2$-Zy4, $UO_2$-M5® and MOX-Zy4 fuel rods with a maximum burnup of 70 GWd/tU. A good agreement between predicted and measured residual diameters after base irradiation, diameter increase at MP and IP level during ramp testing, height of MP and IP ridges and percentage of dish filling at the end of ramp testing was obtained, generally within the known experimental scatter. The 3D scheme was then used with a refined mesh to trigger stress-strain concentration at the triple point where I-SCC is known to initiate. Further work will focus on the implementation of an I-SCC model coupled to the clad behavior law to describe crack propagation at the triple point during power ramps.

## ACKNOWLEDGEMENT

The authors would like to thank EDF and AREVA for their financial and technical support to this research.